# From Noise to Signal: Improving Security Log Anomaly Detection Using LLMs with Endpoint-Specific Logs

Christopher Henshaw[*] and Gour Karmakar[†]

[*] Institute of Innovation, Science and Sustainability
[†]Institute of Innovation, Science and Sustainability, and Centre for Smart Analytics
Email: [*]christopherhenshaw@students.federation.edu.au, [†]Gour.Karmakar@federation.edu.au
Federation University Australia

## Abstract

Cybersecurity remains a significant concern for business, industrial, and government organisations as increasingly complex digital environments create new challenges for protecting systems and information from cyberattacks. The rapid development of Generative Artificial Intelligence (GenAI) has also introduced new opportunities for cybersecurity research, including the use of Large Language Models (LLMs) to support security monitoring and anomaly detection. One important area is the detection of anomalous authentication activity within access control systems. Existing approaches such as Wazuh rely primarily on predefined detection rules, while statistical anomaly detection approaches such as OpenSearch identify deviations from previously observed behavioural patterns. Recent research has investigated LLMs for log anomaly detection because of their ability to interpret semantic and contextual information. However, existing approaches can be affected by prompt construction, noisy log data, and reliance on generic datasets that may lack endpoint-specific authentication behaviours.

To address these limitations, this study develops a standardised instruction-based LLM classification framework for detecting anomalous authentication behaviours, including borderline cases where the distinction between normal and anomalous activity is less obvious. A controlled cybersecurity testbed was developed to generate endpoint-specific authentication data, producing a curated dataset of 75 normal, borderline, and anomalous behavioural scenarios. Three instruction-tuned LLMs, Meta Llama 3.1 8B Instruct, Qwen 2.5 7B Instruct, and GPT-OSS 20B, were evaluated against Wazuh rule-based detection and OpenSearch Anomaly Detection using a common ground-truth severity framework.

Meta Llama 3.1 8B Instruct achieved the strongest overall end-to-end detection performance, with an accuracy of 89.3%, recall of 88.2%, F1-score of 91.8%, and false negative rate of 11.8%. In comparison, Wazuh achieved an accuracy of 52.0% and false negative rate of 68.6%, while OpenSearch achieved an accuracy of 49.3% and false negative rate of 74.5%. Meta Llama also detected 80% of the severity-2 borderline anomalous scenarios, compared with 20% for Wazuh and 15% for OpenSearch. Qwen achieved lower overall detection performance than Meta Llama but recorded the lowest average inference latency and 100% structured-response validity. GPT-OSS demonstrated strong classification performance when valid responses were produced; however, inconsistent structured-output generation substantially reduced its end-to-end reliability across repeated evaluations.

The findings demonstrate the potential of instruction-tuned LLMs to complement existing rule-based and statistical security monitoring approaches by providing an additional semantic analysis layer for authentication behaviour. The developed testbed and evaluation framework also provide a foundation for further investigation using larger datasets, additional endpoint environments, and broader categories of cyberattack behaviour.

Keywords:

authentication anomaly detection; large language models; Wazuh; OpenSearch; security operations; structured output

## 1. Introduction

The rapid growth of computer networks, cloud services, and online platforms has increased the amount of automated activity occurring across modern systems. This includes legitimate activity from monitoring services, scheduled processes, and other automated tools, but also malicious activity such as credential attacks, automated scanning, scraping, and other forms of abuse. For Security Operations Centres (SOCs), distinguishing legitimate activity from suspicious behaviour is an important part of monitoring and protecting networked systems (Eckhoff et al., 2025; Singh et al., 2025).

Cybersecurity monitoring platforms such as Wazuh and OpenSearch use different methods to identify suspicious activity. Wazuh primarily uses predefined rules and conditions to identify known security events, while OpenSearch Anomaly Detection uses statistical methods to identify activity that differs from previously observed behaviour (OpenSearch Project, 2026; Wazuh, Inc., 2024). These approaches provide useful detection capabilities, but they also have limitations. Rule-based detection depends on activity satisfying an existing rule, while statistical anomaly detection depends on identifying a sufficient change from an established baseline (Guide et al., 2024; Zha et al., 2023).

This becomes more challenging when suspicious behaviour is less obvious or continues over time. Repeated authentication attempts, unusual timing patterns, highly consistent interactions, rotating IP addresses, or other automated behaviours may be suspicious when several characteristics are considered together. However, individual events may not always satisfy a predefined detection rule or produce a large enough statistical change to be identified as anomalous. This can make borderline behaviours particularly difficult to detect using a single detection method (Goyal et al., 2023; Muramatsu & Aritsugi, 2024).

Recent research has investigated Large Language Models (LLMs) as another approach to log anomaly detection because of their ability to process text and use information contained within log messages and sequences. Existing studies have used several methods, including prompt-based approaches (Qi et al., 2023; Patel, 2026) and fine-tuned LLMs (Song et al., 2025). These studies demonstrate the potential of LLMs for log anomaly detection, but they also introduce limitations. Prompt-based performance can depend on the information and examples provided to the model, while fine-tuned models depend on factors such as the training data and fine-tuning method. Much of the existing research has also relied on established public log datasets rather than endpoint-specific authentication behaviours generated within a controlled cybersecurity environment.

This study investigates whether recent instruction-tuned LLMs can provide an additional method for detecting suspicious authentication behaviour. Rather than replacing existing detection systems, the LLMs are evaluated as an additional analysis layer that can consider the broader context of a behavioural scenario. The study therefore addresses the following research question:

**How do recent instruction-tuned LLMs perform on endpoint-specific authentication behaviours generated in a controlled cybersecurity environment compared with Wazuh and OpenSearch Anomaly Detection?**

To answer this question, a controlled cybersecurity testbed was developed to generate endpoint-specific authentication activity containing normal, borderline, and anomalous behaviours. A standardised instruction framework was then developed to present these behavioural scenarios to three instruction-tuned LLMs using the same classification requirements, behavioural cues, severity scale, and structured-response format. Their results were compared with Wazuh rule-based detection and OpenSearch Anomaly Detection using the same ground-truth classification framework.

The main contributions of this research are:

- A controlled cybersecurity testbed was developed using a Wazuh server, OpenSearch Anomaly Detection, a monitored Ubuntu endpoint with a Wazuh agent, and a Kali Linux attack host. The testbed was used to generate endpoint-specific authentication activity representing normal, borderline, and anomalous behaviours.

- A standardised instruction framework was developed to provide each LLM with the same behavioural description, seven behavioural cues, six-point severity scale, binary classification threshold, and required JSON response format. An automated evaluation pipeline was used to validate the responses and provide a consistent method for comparing the models.
- Three instruction-tuned LLMs, Meta Llama 3.1 8B Instruct, Qwen 2.5 7B Instruct, and GPT-OSS 20B, were evaluated against Wazuh and OpenSearch. Meta Llama achieved the strongest overall detection performance, with 89.3% accuracy, 88.2% recall, a 91.8% F1-score, and an 11.8% false negative rate. In comparison, Wazuh and OpenSearch achieved accuracies of 52.0% and 49.3% and false negative rates of 68.6% and 74.5%, respectively. GPT-OSS performed well across many of its valid responses, but its low structured-response validity reduced its overall reliability within the evaluation pipeline.
- The evaluation also included 20 severity-2 borderline anomalous scenarios to examine detection performance when suspicious behaviour was less obvious. Meta Llama detected 80.0% of these scenarios, followed by Qwen at 55.0%, Wazuh at 20.0%, and OpenSearch at 15.0%. GPT-OSS produced valid structured responses for only five of the 20 scenarios during the primary evaluation, with all five correctly classified as Anomaly. The remaining 15 scenarios produced invalid responses and were therefore not treated as classifications of Normal.

The remainder of this paper is organised as follows. Section 2 reviews existing research in log and behavioural anomaly detection. Section 3 describes the research methodology, including the experimental testbed, dataset generation, ground-truth labelling, baseline detection systems, LLM classification pipeline, and evaluation framework. Section 4 presents and compares the experimental results. Section 5 discusses the findings, limitations, and directions for future research.

## 2. Related Work

Network security monitoring platforms such as Wazuh and OpenSearch are widely used in Security Operations Centres (SOCs) to detect anomalous behaviour across hosts, including endpoints, networks, and applications (Ahmed et al., 2016; Chandola et al., 2009). These platforms ingest large volumes of telemetry, including system logs, network flow data, authentication events, and application-level activity, and apply a combination of rule-based detection, statistical analysis, and machine-learning techniques to identify deviations from expected behaviour (Amazon Web Services, 2023; Wazuh, Inc., 2024). Wazuh and OpenSearch are presented in the following sections:

### 2.1 Rule-Based Detection

Wazuh primarily uses predefined rules, signatures, and threshold-based detection to identify security events and generate alerts (Wazuh, Inc., 2024). This approach is effective for detecting known attack patterns, authentication failures, policy violations, and other activities that satisfy predefined detection conditions. As illustrated in Figure 1, security events and logs generated by monitored endpoints are collected by Wazuh agents and forwarded to the Wazuh Manager. The manager decodes the received events, evaluates them against predefined detection rules, and generates alerts when relevant rule conditions are satisfied. These alerts are subsequently indexed and stored by the Wazuh Indexer and made available through the Wazuh Dashboard for visualisation, monitoring, and investigation. However, the effectiveness of this approach depends heavily on the coverage and configuration of its detection rules. Behaviours that do not satisfy existing rule conditions, or that require broader contextual interpretation across multiple events, may therefore be more difficult to identify (Agarwal & Hussain, 2018; Guide et al., 2024).

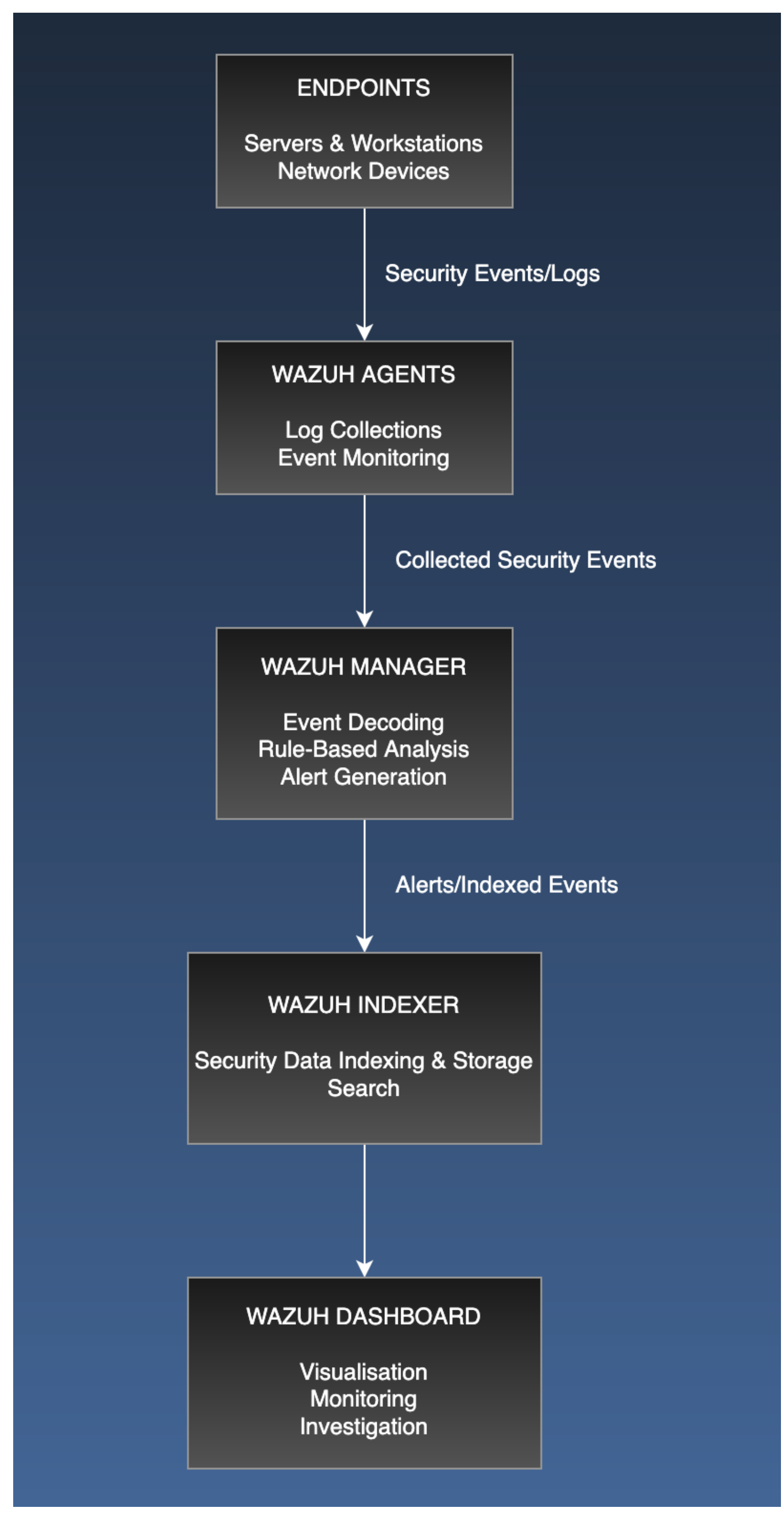

*Figure 1. Conceptual architecture of Wazuh rule-based security event detection and analysis.*

## 2.2 Statistical Anomaly Detection

OpenSearch provides an anomaly detection framework based on unsupervised machine learning for identifying unusual changes within time-series data. The OpenSearch Anomaly Detection plugin uses the Random Cut Forest (RCF) algorithm, which models incoming data streams and evaluates new observations according to how strongly they deviate from previously observed behaviour (Guha et al., 2016; OpenSearch Project, 2026). Unlike clustering algorithms that organise observations into groups based on similarity, RCF is designed to identify anomalous observations within dynamic data streams.

OpenSearch applies RCF to one or more configured features derived from the monitored data. These features can represent aggregated measurements such as event counts, averages, minimum or maximum values, or other numerical characteristics observed over defined time intervals. As new observations are processed, the detector produces an anomaly grade and confidence score that indicates whether the observed behaviour represents an unusual deviation from the modelled data stream (OpenSearch Project, 2026).

Within cybersecurity monitoring environments, this approach enables OpenSearch to identify statistical changes in security telemetry without requiring predefined attack signatures. For authentication data, for example, a detector may monitor features such as the number of failed authentication attempts within a defined interval. A substantial deviation from the previously observed pattern may receive an elevated anomaly grade, allowing unusual authentication activity to be identified even when no predefined security rule has been triggered.

The effectiveness of this approach, however, depends on the selected features, the quality and continuity of the input data, and the behavioural patterns represented by the model. OpenSearch notes that feature selection and data quality can affect anomaly detection performance, including precision and recall. Consequently, statistical anomaly detection is effective for identifying deviations from observed behaviour but does not directly determine whether the underlying behaviour is semantically plausible or malicious. This distinction becomes particularly important when anomalous activity persists over time and motivates the comparative analysis presented in the following section.

## 2.3 The Effectiveness of Wazuh and OpenSearch for Anomaly Detection

Wazuh and OpenSearch provide complementary approaches to security monitoring, but their detection capabilities are based on fundamentally different mechanisms. Wazuh identifies security events primarily through predefined rules and detection conditions, whereas OpenSearch Anomaly Detection uses unsupervised machine learning to identify statistical deviations from previously observed behaviour. These approaches are effective for detecting known security events and measurable deviations; however, their effectiveness depends on whether suspicious behaviour satisfies a predefined rule or produces a sufficient statistical deviation.

For a rule-based system such as Wazuh, detection performance depends heavily on the coverage, configuration, and thresholds of the available rules. Events that satisfy known detection conditions, such as repeated authentication failures or attempts to authenticate using invalid accounts, can be identified reliably. However, suspicious activity that remains below configured thresholds or consists of individually legitimate events may not generate an alert. This limitation becomes particularly relevant when malicious activity is distributed across longer periods or performed using otherwise legitimate operations, as individual events may provide insufficient evidence of malicious behaviour when analysed independently.

Statistical anomaly detection introduces a different challenge. As discussed in Section 2.2, OpenSearch Anomaly Detection uses Random Cut Forest (RCF) to identify observations that deviate from patterns represented within streaming data. Changes in the underlying data distribution over time can affect anomaly-detection performance, a problem commonly associated with concept drift (Gama et al., 2014). The effectiveness of statistical anomaly detection depends not only on the detection algorithm but also on the selected features, quality of the input data, and continued relevance of the behavioural patterns represented by the model.

These limitations are particularly important in cybersecurity because adversarial behaviour does not always produce abrupt or high-volume deviations. Attackers may attempt to reduce their visibility by distributing activity over time, limiting the frequency or volume of malicious actions, or including behaviours that resemble normal activity. Such low-rate and mimicry-based strategies can make malicious activity more difficult to distinguish from normal behaviour using individual events or statistical deviation alone (Goyal et al., 2023; Muramatsu & Aritsugi, 2024). Rather than producing a single obvious anomaly, the indication of malicious activity may emerge from the relationship between multiple events, their timing, repetition, and operational context.

These limitations do not diminish the operational value of Wazuh or OpenSearch, but instead, they demonstrate that rule-based and statistical detection approaches capture different dimensions of anomalous behaviour. This creates an opportunity for a complementary detection mechanism capable of evaluating behavioural context in addition to rule matches and statistical deviation. The following sections therefore examine machine-learning and Large Language Model (LLM)-based approaches, with particular attention to whether semantic and contextual reasoning can provide additional information for security anomaly detection.

### 2.4 Machine Learning and Deep Learning for Anomaly Detection

Machine learning (ML) and deep learning (DL) techniques have been widely investigated for anomaly and intrusion detection in cybersecurity environments. Unlike predefined rule-based approaches, ML-based detection systems learn patterns from historical data and use these patterns to classify new observations or identify deviations that may indicate malicious behaviour (Ahmed et al., 2016; Chandola et al., 2009). These techniques have been applied to a range of security telemetry, including network traffic, authentication activity, system logs, and user behaviour.

Traditional ML approaches used for anomaly and intrusion detection include decision trees, Random Forests, Support Vector Machines (SVMs), k-nearest neighbours, clustering methods, and anomaly-specific algorithms such as Isolation Forest. More recent research has incorporated deep learning architectures, including Convolutional Neural Networks (CNNs), autoencoders, Recurrent Neural Networks (RNNs), and Long Short-Term Memory (LSTM) networks (Landauer et al., 2022; Le & Zhang, 2022). These models can identify complex relationships within large datasets, while architectures such as LSTM are particularly useful for analysing sequential and time-dependent security data (Ali et al., 2024).

Transformer-based architectures have also been investigated for analysing sequential cybersecurity data. Their attention mechanisms allow relationships between events to be identified across longer sequences. This has led to the use of Transformer and hybrid architectures for intrusion and anomaly detection. However, increased model complexity can also introduce greater computational requirements, training costs, and interpretability challenges. These limitations may affect their suitability for resource-constrained or real-time cybersecurity environments.

Despite their detection capabilities, ML and DL approaches remain dependent on the quality and representativeness of the data used during model development. Supervised approaches require sufficiently labelled examples of normal and malicious behaviour, while unsupervised approaches attempt to identify patterns without explicit attack labels. In practical cybersecurity environments, obtaining large and accurately labelled datasets can be difficult, particularly for new or infrequently observed attacks. Differences between benchmark datasets and real-world environments may also limit how well a model performs outside the conditions in which it was originally trained (Le & Zhang, 2022).

A further challenge is concept drift, where patterns of network activity change over time as users, systems, applications, and adversarial behaviours evolve. Models trained on historical data may therefore experience reduced detection performance as the operational environment changes. Recent research continues to identify concept and feature drift as important challenges for ML-based intrusion detection, creating a need for adaptive models and continued monitoring of model performance (Gama et al., 2014; Shyaa et al., 2024).

Interpretability is another important consideration when deploying complex ML and DL models in cybersecurity environments. Models such as deep neural networks, LSTMs, and Transformers can provide effective classifications while providing limited information about why an individual prediction was made. Explainable Artificial Intelligence (XAI) techniques, including SHAP and LIME, can partially address this limitation by providing information about the features that influence model predictions (Lundberg & Lee, 2017; Ribeiro et al., 2016). These techniques can provide security analysts with additional information when interpreting classification decisions. However, they generally provide an additional explanation layer over the underlying model rather than directly reasoning about the broader behavioural context.

Overall, ML and DL approaches provide capabilities beyond static rule-based detection by learning complex patterns from security data. However, their effectiveness can still be influenced by training data, feature selection, changing network behaviour, computational requirements, and model interpretability. These challenges support further investigation into detection approaches that can incorporate broader contextual information and provide understandable assessments of suspicious behaviour.

### 2.5 Large Language Models for Anomaly Detection

Large Language Models (LLMs) have recently been investigated for log-based anomaly detection because of their ability to process and interpret text-based information. System and security logs can contain both structured and unstructured information describing events, errors, authentication activity, and system behaviour. Unlike approaches that rely mainly on numerical features or predefined rules, LLMs can use the information contained within log messages and sequences to help identify unusual behaviour.

Qi et al. (2023) investigated the use of ChatGPT for log anomaly detection using log preprocessing and prompt construction. Their prompt included a task description, domain knowledge, and a required response format. An important part of their approach was the inclusion of human domain knowledge to help the model interpret the logs. However, providing examples of normal or anomalous logs within the prompt may influence how the model interprets other log patterns and could make it more difficult to generalise beyond the examples provided.

Other studies have explored different ways of using language models for log anomaly detection. Han et al. (2023) proposed LogGPT, which uses a GPT-based model to learn normal log sequences and identify events that do not follow the expected sequence. The model was evaluated using three public log datasets and achieved improved anomaly detection compared with the baseline approaches included in the study. Guan et al. (2024) later proposed LogLLM, which combines BERT to extract information from individual log messages with Llama to classify sequences of logs. Their approach was evaluated across four public datasets and showed that information contained within log messages and sequences can be used by LLM-based models to identify anomalous behaviour.

Hadadi et al. (2024) examined how LLM-based anomaly detection performs when log data changes over time. The study evaluated GPT models using unstable logs generated from different versions of the same software system. A fine-tuned GPT-3 model performed slightly better than the supervised approaches used for comparison, with the difference becoming more noticeable as the log sequences changed. However, the authors also noted that the practical improvement was not clear in every case. This shows that LLMs may provide advantages when log patterns change, but further testing across different datasets and environments is still required.

Song et al. (2025) investigated LLM fine-tuning for insider-threat detection using employee behaviour logs. Their approach used contrastive learning and evaluated three different fine-tuning strategies using data from tested and non-tested users. The first used training data from non-tested users, the second included data from both user groups, and the third used a two-stage fine-tuning process to balance generalisation with user-specific behavioural information. The final model was used to detect insider threats using temporal web, file, email, device, and other behavioural information from the CMU CERT v6.2 insider-threat dataset. However, the study focused specifically on employee behaviour within an established public dataset. Its performance also depended on the selected data-augmentation methods and threshold values used to construct the behavioural representations.

LLMs have also been combined with other detection techniques rather than being used as standalone detectors. LogRAG, for example, combines log anomaly detection with Retrieval-Augmented Generation (RAG), allowing an LLM to reassess logs that have already been identified as anomalous by another detection component (Zhang et al., 2024). This provides an example of how an LLM can be used as an additional analysis layer within an existing detection pipeline rather than replacing the original detection system.

Patel (2026) compared traditional machine learning approaches, including SVM and Random Forest, fine-tuned transformer models such as BERT and RoBERTa, and prompt-based LLMs including GPT-3.5, GPT-4, and LLaMA-3. The evaluation used public datasets including HDFS, BGL, Thunderbird, and Spirit. The prompt included contextual information about the log-generating system and timestamps, keywords associated with anomalous activity, and a required response format. The zero-shot and five-shot results showed that adding contextual information to the prompt improved GPT-4's detection performance compared with the relevant baseline configurations. However, including specific anomaly-related keywords may also influence how the model classifies conflicting, noisy, or borderline logs. The contextual information used in the study was also general and did not include the endpoint-specific authentication behaviours examined in this research.

Despite these advantages, using LLMs for anomaly detection also introduces several challenges. Detection performance can depend on the model, prompt design, training or fine-tuning method, and how the log information is presented. Larger models may also require more computational resources and introduce additional inference delays. These factors become important in cybersecurity monitoring environments where large numbers of events may need to be analysed continuously.

Overall, existing research shows that LLMs provide another way of approaching log anomaly detection by using information contained within individual log messages and sequences of events. However, much of the existing research has relied on established public datasets or specialised log-analysis frameworks. There is still limited research comparing recent instruction-tuned LLMs directly with established detection approaches such as Wazuh and OpenSearch using endpoint-specific authentication behaviours. This provides the basis for the experimental approach evaluated in this study.

### 2.6 Research Gap Summary

The review of existing approaches highlights a fundamental limitation in current online agent anomaly detection methods used in network security monitoring. Rule-based systems and statistical baselining techniques, such as those employed by Wazuh and OpenSearch, are effective at identifying abrupt deviations and short-lived anomalies but may lose sensitivity when anomalous behaviour persists over time due to baseline normalisation and concept drift (Gama et al., 2014). Similarly, traditional machine learning approaches rely heavily on historical data distributions and engineered features, making them susceptible to concept drift and baseline normalisation in dynamic network environments.

In practical network operations, many automated agents are intentionally designed to operate persistently, quietly, and within stable behavioural patterns (Stringhini et al., 2013). Rather than generating sharp deviations, these agents exhibit consistent activity over extended periods, often mimicking legitimate usage characteristics. Research on mimicry attacks shows that malicious behaviour can evade anomaly detectors by resembling benign operations (Goyal et al., 2023; Muramatsu & Aritsugi, 2024).

Although such behaviour may become statistically normal according to learned models, it remains operationally implausible for human users when evaluated in terms of temporal regularity, continuity, and behavioural constraints. Existing detection approaches lack the ability to explicitly reason about these higher-level characteristics. As network environments grow in scale and complexity, there is an increasing need for detection mechanisms that can provide interpretable and context-aware assessments of anomalous behaviour.

Prior studies in log anomaly detection have commonly used established public benchmark datasets such as HDFS, BGL, Thunderbird, and Spirit (Wu et al., 2023; Patel, 2026). These benchmark datasets represent system logs collected from distributed computing and large-scale computing environments. For example, HDFS contains logs from the Hadoop Distributed File System, while BGL, Thunderbird, and Spirit originate from large-scale supercomputing systems (Wu et al., 2023; Oliner & Stearley, 2007). Although these datasets remain widely used for log anomaly detection research, they do not specifically represent endpoint authentication behaviours generated within a contemporary cybersecurity monitoring environment. In addition, limited research has directly compared recent instruction-tuned LLMs with operational detection approaches such as Wazuh and OpenSearch using the same endpoint-specific behavioural scenarios. This gap motivates a controlled comparison using newly generated and labelled authentication behaviours containing normal, borderline, and anomalous scenarios.

## 3. Research Methodology

A key contribution of this research is the development of a controlled cybersecurity testbed for generating endpoint-specific authentication behaviours. The testbed was designed to generate labelled normal, borderline, and anomalous authentication activity within a reproducible environment. This allowed the same behavioural scenarios to be evaluated using rule-based detection, statistical anomaly detection, and LLM-based semantic classification.

Another important aspect of the research is the use of a standardised instruction framework for the LLM-based evaluation. Each model was provided with the same behavioural descriptions, classification requirements, severity scale, behavioural cues, and structured-response format. This provided a consistent method for comparing the selected LLMs and reduced differences caused by variations in the instructions provided to each model.

The study also specifically evaluated severity-2 borderline anomalous behaviours generated from the testbed. These scenarios represented suspicious authentication activity located closest to the Normal/Anomaly decision threshold and were used to compare the detection performance of the LLMs, Wazuh, and OpenSearch when the anomalous behaviour was less obvious.

This section describes the experimental testbed, dataset generation and ground-truth labelling process, baseline detection systems, LLM classification pipeline, prompt design, model selection, and evaluation framework used throughout the experiment.

### 3.1 Experimental Testbed

To evaluate the proposed semantic detection framework, a controlled experimental testbed was developed to generate, collect, and analyse authentication-based security events. The purpose of the testbed was to collect relevant log data within a reproducible environment where normal user activity and simulated adversarial behaviour could be evaluated across traditional detection systems and LLM-based classification.

As illustrated in Figure 2, the environment was hosted on an Acer Predator Helios Neo 16 laptop running Ubuntu 24.04.4 LTS. The system contained an Intel Core i9-14900HX processor, 32 GB RAM, and an NVIDIA RTX 4060 GPU. VirtualBox was used to create an isolated virtual network environment consisting of a Wazuh monitoring server, a monitored Linux endpoint, and a Kali Linux attacker machine.

The Wazuh monitoring environment was deployed using Ubuntu Server 22.04.5 LTS with Wazuh Server version 4.14.2-rc4. The Wazuh Indexer was based on OpenSearch 2.19.4 and included the OpenSearch Anomaly Detection plugin, which was used as the statistical detection baseline. The monitored endpoint consisted of an Ubuntu Desktop 22.04.5 LTS virtual machine configured as a Wazuh agent. A Kali Linux Rolling 2025.4 virtual machine was used to generate controlled adversarial activity, including SSH authentication attempts and suspicious access behaviours.

Authentication events generated on the monitored endpoint were recorded in the Linux authentication log at /var/log/auth.log. As shown in Figure 2, these events were collected by the Wazuh agent and processed by the Wazuh monitoring environment to generate security alerts. The authentication activity was also used to construct behavioural descriptions for LLM-based analysis. The testbed therefore generated both legitimate administrative activity and simulated adversarial scenarios, allowing rule-based detection, statistical anomaly detection, and semantic LLM-based classification to be evaluated using the same behavioural environment.

The LLM evaluation component was executed separately using Google Colab. Behavioural scenarios generated from the experimental environment were submitted to three instruction-tuned LLMs through API-based inference. In this context, instruction-tuned refers to models that have been trained to follow natural-language instructions and perform specified tasks. Each model was provided with the same standardised classification instructions rather than being additionally fine-tuned as part of this study. Separating the LLM evaluation pipeline from the testbed allowed the traditional detection environment and semantic analysis component to operate independently, representing a potential deployment model in which LLM-based analysis functions as an additional layer within an existing security monitoring workflow.

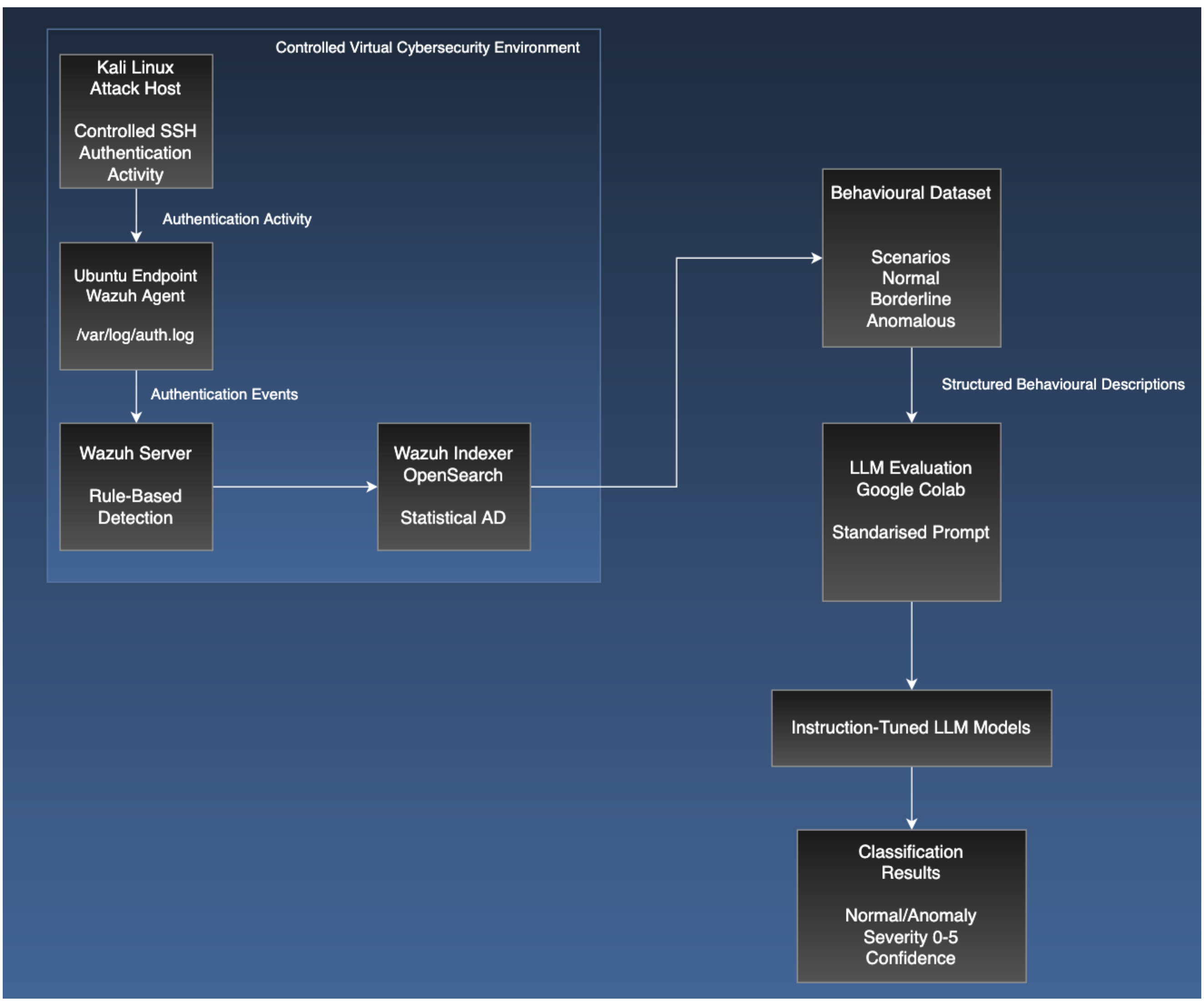

*Figure 2. Experimental testbed architecture illustrating authentication event generation, Wazuh/OpenSearch processing, behavioural extraction, and the model-agnostic LLM-based semantic analysis framework.*

### 3.2 Dataset Generation and Collection

Authentication-based security events were generated within the controlled virtual environment described in Section 3.1. The objective was to create a dataset containing both expected user activity and behaviours representative of suspicious or automated activity commonly observed within cybersecurity environments (Buczak & Guven, 2016).

The Ubuntu endpoint acted as the monitored system, while the Kali Linux virtual machine was used to generate controlled adversarial behaviours. Activity was captured through Linux authentication logs (/var/log/auth.log) and processed through the Wazuh monitoring environment. Generated events included successful authentication sessions, failed login attempts, invalid user authentication attempts, and repeated access behaviours designed to simulate brute-force style activity.

The raw authentication events generated during testing were analysed and converted into a structured behavioural dataset for evaluation. Rather than evaluating individual log entries in isolation, related authentication activity was represented as contextual behavioural scenarios describing authentication patterns and user activity. For example, one borderline scenario described "Multiple failed SSH login attempts followed by successful login from same IP." This scenario was assigned an expected anomaly severity score of 2 because the combination and sequence of authentication events represented suspicious behaviour requiring investigation. Representing the activity in this form allowed the LLMs to evaluate the broader behavioural context rather than relying only on individual authentication events.

The experimental environment generated hundreds of raw authentication events, which were subsequently analysed and consolidated into a curated evaluation dataset consisting of 75 representative behavioural scenarios used for comparative evaluation. The final dataset contained 20 normal, 24 borderline or suspicious, and 31 anomalous behavioural scenarios, as shown in Table 1. The dataset intentionally included a higher proportion of suspicious and anomalous behaviours to evaluate detection sensitivity and false-negative behaviour rather than reproduce the class distribution of normal enterprise network traffic.

*Table 1: Behavioural Dataset Distribution*

| Behavioural Category | Number of Scenarios | Percentage |
|---|---|---|
| Normal Behaviour | 20 | 26.7% |
| Borderline/Suspicious Behaviour | 24 | 32.0% |
| Anomalous Behaviour | 31 | 41.3% |
| Total | 75 | 100% |

The inclusion of borderline behaviours was designed to evaluate scenarios where detection decisions were less obvious. Of the 24 borderline scenarios, four were assigned severity 1 and were therefore treated as Normal during binary evaluation, while 20 were assigned severity 2 and treated as Anomaly. This distinction is based on the ground-truth severity threshold described in Section 3.3. These cases were included to assess behaviours that may not trigger traditional rule-based or statistical detection but could indicate suspicious behaviour when broader contextual information is considered (Blázquez-García et al., 2021).

### 3.3 Ground Truth Labelling Process

To provide a consistent reference for evaluating detection performance, each behavioural scenario was assigned a ground-truth anomaly severity score before the detection results were evaluated. The ground-truth labels represented the expected classification of each scenario and were used as the common reference for comparing Wazuh rule-based detection, OpenSearch statistical anomaly detection, and LLM-based classification. Establishing the ground truth before evaluating detector outputs ensured that the expected classifications were not determined from, or altered in response to, the predictions produced by the evaluated detection approaches.

A six-point severity scale ranging from 0 to 5 was used to represent increasing levels of suspicious behaviour. Labels were assigned based on observable behavioural characteristics, including authentication success or failure, repeated access attempts, timing patterns, activity consistency, and indicators associated with automated behaviour (Buczak & Guven, 2016). The severity scale was designed to distinguish clearly legitimate behaviour from increasingly suspicious or malicious activity while retaining borderline cases where the distinction between Normal and Anomaly was less obvious. Table 2 presents the ground-truth severity scale and corresponding evaluation classifications.

*Table 2: Ground Truth Severity Scale*

| Severity Score | Description | Evaluation Classification |
|---|---|---|
| 0 | Expected human or legitimate system behaviour | Normal |
| 1 | Slightly unusual behaviour but still considered plausible | Normal |
| 2 | Suspicious behaviour requiring investigation | Anomaly |
| 3 | Behaviour strongly suggesting automation or misuse | Anomaly |
| 4 | High-confidence indicator of malicious or automated activity | Anomaly |
| 5 | Critical behaviour representing a high-confidence attack pattern | Anomaly |

For comparative evaluation, the six-point severity scale was converted into a binary classification using a predefined decision threshold. Behavioural scenarios assigned severity scores of 0 or 1 were classified as Normal, while scenarios assigned severity scores from 2 to 5 were classified as Anomaly. The same threshold was applied consistently throughout the evaluation so that outputs from the different detection approaches could be compared using a common binary classification framework. The threshold placed severity 2 behaviours within the Anomaly class because these scenarios represented suspicious activity requiring investigation, supporting the study's emphasis on reducing false negatives within cybersecurity monitoring environments.

The application of this threshold resulted in 24 Normal scenarios and 51 Anomaly scenarios across the complete scenario evaluation dataset. As described in Section 3.2, the 24 borderline or suspicious scenarios contained four severity-1 scenarios and 20 severity-2 scenarios. Consequently, the four severity-1 borderline scenarios were included in the Normal class, while the 20 severity-2 borderline scenarios were included in the Anomaly class during binary evaluation. This distinction allowed borderline behaviours to remain part of the dataset while preserving a consistent ground-truth decision rule.

For the calculation of classification metrics, Anomaly was defined as the positive class and Normal as the negative class. A correctly detected anomalous scenario was therefore recorded as a true positive (TP), while an anomalous scenario classified as Normal was recorded as a false negative (FN). Similarly, a correctly identified Normal scenario was recorded as a true negative (TN), while a Normal scenario classified as Anomaly was recorded as a false positive (FP). These definitions were applied consistently to Wazuh, OpenSearch, and the evaluated LLMs and form the basis of the evaluation metrics described later in this section.

The inclusion of multiple severity levels allowed each detection approach to be evaluated across clearly legitimate behaviour, borderline activity, and more obvious anomalous or malicious behaviour. This provided a consistent basis for assessing whether LLM-based semantic analysis could identify suspicious behavioural patterns that may not directly satisfy predefined rule conditions or produce sufficiently strong statistical deviations.

### 3.4 Baseline Detection Systems

Two existing cybersecurity detection approaches were selected as baseline systems for comparison against LLM-based classification: Wazuh rule-based detection and OpenSearch Anomaly Detection. These systems represent two common approaches used within security monitoring environments: predefined rule matching and statistical anomaly detection (OpenSearch Project, 2026; Wazuh, Inc., 2024).

The purpose of including these baselines was to establish representative rule-based and statistical detection approaches against which the proposed LLM-based semantic analysis could be evaluated. Each approach analysed the authentication behaviour according to its own detection mechanism, with the resulting detections subsequently compared against the common ground-truth classification described in Section 3.3.

#### *3.4.1 Wazuh Rule-Based Detection*

Wazuh Server version 4.14.2-rc4 was configured as the rule-based detection baseline. Authentication events generated within the test environment were collected from Linux authentication logs (/var/log/auth.log) and analysed using Wazuh's predefined SSH and PAM detection rules (Wazuh, Inc., 2024).

The monitored Ubuntu endpoint was configured as a Wazuh agent, while the Kali Linux system generated controlled authentication activity. Wazuh processed these events and generated alerts when observed behaviour matched predefined rule conditions. Alert generation therefore depended on predefined detection logic rather than semantic interpretation of behavioural context (Buczak & Guven, 2016). The generated alerts included both successful administrative activity and failed authentication attempts, allowing the observed Wazuh detections to be compared with the ground-truth classifications used in the evaluation.

The primary SSH attack-related rule observed during testing was Rule 5710, which detects attempts to authenticate using non-existent SSH users. This activity relates to MITRE ATT&CK techniques T1110.001 (Password Guessing) and T1021.004 (SSH), which represent password-guessing and remote SSH activity, respectively (MITRE, 2026). Table 3 presents the Wazuh rule ID, rule level, description, and number of alerts observed during testing.

*Table 3: Wazuh Authentication Rules Observed During Testing*

| Rule ID | Wazuh Rule Level | Description | Alert Count |
|---|---|---|---|
| *5710* | 5 | sshd: Attempt to login using a non-existing user | 644 |
| *5501* | 3 | PAM: Login session opened | 247 |
| *5502* | 3 | PAM: Login session closed | 214 |
| *5402* | 3 | Successful sudo to ROOT executed | 152 |
| *5715* | 3 | sshd: authentication success | 54 |
| *5503* | 5 | PAM: User login failed | 6 |

While Wazuh provides effective identification of security events that satisfy known detection conditions, its effectiveness depends on the coverage and configuration of the available rules. Behaviours that do not satisfy existing rule conditions, or that require broader contextual interpretation across multiple authentication events, may therefore remain undetected (Buczak & Guven, 2016; Chandola et al., 2009).

#### *3.4.2 OpenSearch Anomaly Detection*

OpenSearch Anomaly Detection was selected as the statistical anomaly detection baseline. The experimental environment used Wazuh Indexer based on OpenSearch version 2.19.4 with the OpenSearch Anomaly Detection plugin enabled (OpenSearch Project, 2026).

Unlike rule-based detection approaches, OpenSearch identifies unusual activity by evaluating statistical deviations within time-series data. As described in Section 2.2, the Anomaly Detection plugin uses the Random Cut Forest (RCF) algorithm to model incoming data and identify observations that deviate from previously observed behaviour (Guha et al., 2016; OpenSearch Project, 2026). Authentication activity collected from the monitored environment was analysed using configured numerical features derived from the authentication data, allowing statistical changes in behaviour to be evaluated without relying on predefined attack signatures.

The OpenSearch baseline was used to evaluate whether statistical changes in authentication behaviour could identify suspicious activity represented within the evaluation dataset. OpenSearch detections were subsequently mapped to the same binary Normal and Anomaly framework described in Section 3.3, allowing its performance to be compared with Wazuh and the evaluated LLMs using a common ground-truth reference.

Although statistical anomaly detection can identify unexpected changes in behaviour, its effectiveness depends on factors including the selected features and the behavioural patterns represented by the model. Changes in the underlying data distributed over time can also affect anomaly-detection performance through concept drift (Gama et al., 2014). Consequently, behaviour that produces insufficient deviation may be more difficult to identify despite remaining suspicious when considered within its broader behavioural context.

### **3.5 LLM Classification Pipeline**

The LLM classification pipeline was developed to evaluate whether semantic interpretation of authentication behaviours could improve anomaly detection when compared with traditional rule-based and statistical approaches. Rather than analysing individual log entries directly, each behavioural scenario was converted into a structured natural language description and presented to multiple instruction-tuned LLMs for evaluation (Qi et al., 2023; Xu et al., 2024).

Three instruction-tuned LLMs were selected for comparison: Meta Llama 3.1 8B Instruct, Qwen 2.5 7B Instruct, and GPT-OSS 20B. Each model received the same behaviour description and identical classification instructions to maintain consistent evaluation conditions. A Python-based evaluation program was developed to read the behavioural dataset, construct the standardised prompt for each scenario, submit requests to the selected LLMs, validate the returned responses, record inference latency, and compare valid predictions against the labelled ground truth.

Model inference was performed using the Hugging Face Router API through an OpenAI-compatible interface. A temperature of 0.0 was used to reduce response variability, and the maximum response length was limited to 300 tokens. The same generation settings were applied across all three models to maintain consistency throughout the evaluation.

Each prompt instructed the model to determine whether the observed behaviour represented normal activity or an anomaly, assign an anomaly severity score between 0 and 5, provide a confidence score between 0.0 and 1.0, identify the behavioural indicators that influenced the decision, and provide a brief rationale supporting the classification. Responses were required to conform to a predefined JSON structure containing classification, anomaly grade, confidence score, identified behavioural signals, and supporting rationale. The prompt design and behavioural indicators are described further in Section 3.6.

Returned responses were automatically validated by the program before being accepted for analysis. Validation confirmed that the required JSON structure and fields were present, the anomaly grade was within the permitted range, the confidence value was valid, and the returned classification was consistent with the predefined severity decision rule. Responses that failed these requirements were recorded as invalid responses rather than accepted as successful classifications.

For valid responses, each model's predicted anomaly grade was converted into a binary classification using the same decision threshold defined for the ground truth in Section 3.3. Predicted grades of 0 or 1 were classified as Normal, while grades 2 to 5 were classified as Anomaly. The resulting predictions were then compared with the ground-truth classifications to determine detection performance. The pipeline also recorded inference latency and response validity to support the operational evaluation presented in Section 4.

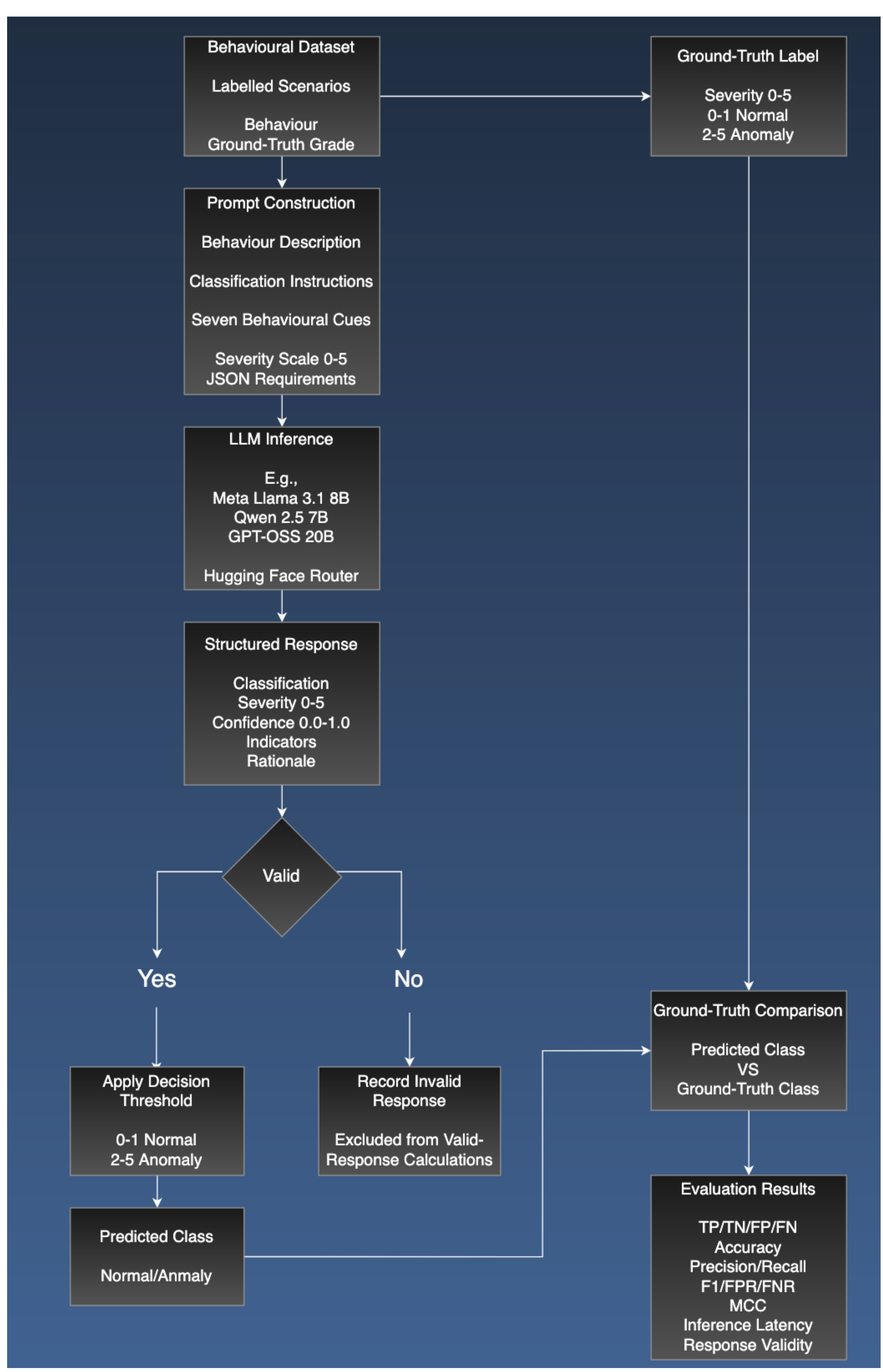

*Figure 3. LLM classification pipeline showing the transformation of authentication behaviours into structured prompts, model inference, response validation, and binary anomaly classification.*

### 3.6 Prompt Engineering and Behavioural Cues

A standardised prompt template was developed to ensure that each LLM evaluated behavioural patterns using identical instructions and decision criteria. Using the same prompt structure across all evaluated models reduced experimental variation and allowed differences in classification performance to be attributed more consistently to model behaviour rather than differences in the instructions provided (White et al., 2023).

Each behavioural scenario was presented as a natural-language description of authentication activity rather than as raw system logs. This approach was selected to evaluate whether instruction-tuned LLMs could interpret the broader behavioural context of authentication activity and identify suspicious patterns using semantic information rather than relying solely on individual log messages, predefined signatures, or statistical deviations (Guan et al., 2024; Qi et al., 2023).

For each behavioural scenario, the prompt instructed the model to:

1. Classify the behaviour as either Normal or Anomaly;
2. Assign an anomaly severity score between 0 and 5;
3. Provide a confidence score between 0.0 and 1.0;
4. Identify the behavioural indicators that influenced the classification; and
5. Provide a brief rationale supporting the decision.

The prompt used the severity definitions established by the ground-truth framework in Section 3.3 so that all models interpreted the classification scale consistently. In addition to these classification instructions, seven behavioural cues were provided to guide the model's assessment:

- Circadian breaks: whether activity showed expected periods of inactivity associated with normal human behaviour;
- Timing variance: whether the timing between actions showed natural variation or unusual consistent intervals;
- Impossible travel or inconsistent geolocation: whether authentication activity occurred across locations that would be implausible within the observed time period;
- Repetitive sequences: whether the same sequence of actions was repeatedly performed with little or no variation;
- Rate-limit ignoring or perfectly timed retries: whether repeated actions continued despite rate limiting or occurred at unusually precise intervals;
- Rotating IP addresses with identical fingerprints: whether source addresses changed while other behavioural or system characteristics remained identical; and
- Extremely fast interactions: whether actions occurred at speeds unlikely to represent normal human interaction.

These cues were provided as general behavioural indicators rather than scenario-specific labels. The prompt did not specify which cues were present within an individual scenario or provide the expected classification. Instead, each model was required to determine whether the available behavioural information provided evidence of anomalous activity. This allowed the same behavioural criteria to guide each model while preserving the classification task itself.

Finally, responses were required to follow the predefined JSON structure described in Section 3.5, allowing the evaluation program to extract the classification, anomaly grade, confidence score, identified behavioural signals, and supporting rationale consistently across all evaluated models. The same prompt template and behavioural cues were used throughout the evaluation.

### 3.7 Model Selection

Three instruction-tuned Large Language Models (LLMs) were selected to evaluate whether semantic reasoning could improve anomaly detection within cybersecurity monitoring environments. The selected models were: Meta Llama 3.1 8B Instruct (Meta AI, 2024), Qwen 2.5 7B Instruct (Qwen Team, 2024), and GPT-OSS 20B (OpenAI, 2025). These models were selected to provide representation from multiple LLM families while remaining suitable for comparative evaluation using the standardised classification pipeline described in Section 3.5.

The selected models differed in model size, architecture, training methodology, and development organisation (Meta AI, 2024; Qwen Team, 2024; OpenAI, 2025). Meta Llama 3.1 8B Instruct and Qwen 2.5 7B Instruct provided similarly sized models from different families, while GPT-OSS 20B provided a larger model for comparison. Evaluating models from multiple families allowed the experiment to examine whether semantic anomaly detection performance was consistent across different LLM implementations rather than reflecting the characteristics of a single model.

Instruction-tuned models were selected because the experiment task required each model to follow a predefined set of classification instructions, interpret behavioural descriptions, and return structured information for automated evaluation. Each model was therefore evaluated using the same prompt template, behavioural cues, generation settings, and validation requirements described in Sections 3.5 and 3.6. This provided consistent experimental conditions while allowing differences in classification performance and response reliability to be observed across the selected models.

The objective of model selection was not to identify the most capable general-purpose LLM, but to determine whether multiple instruction-tuned models could provide useful semantic interpretations of authentication behaviours when compared with traditional rule-based and statistical detection methods. This supports the broader objective of evaluating LLM-based classification as an additional semantic analysis layer within existing Security Information and Event Management (SIEM) environments.

### 3.8 Evaluation Framework and Performance Metrics

A common evaluation framework was used to compare the performance of Wazuh rule-based detection, OpenSearch Anomaly Detection, and the three LLM-based classifiers. Although each approach used a different detection mechanism, their outputs were evaluated against the same ground-truth classifications established in Section 3.3. This provided a consistent basis for comparing rule-based, statistical, and semantic anomaly detection performance.

For each behavioural scenario, the classification produced by each detection approach was compared with its corresponding ground-truth classification. Anomaly was defined as the positive class and Normal as the negative class. A correctly detected anomalous scenario was recorded as a true positive (TP), while an anomalous scenario classified as Normal was recorded as a false negative (FN). A correctly identified Normal scenario was recorded as a true negative (TN), while a Normal scenario classified as Anomaly was recorded as a false positive (FP). These outcomes formed the confusion matrix for each detection approach and provided the basis for calculating the performance metrics.

For the end-to-end LLM evaluation, a valid structured response was required for a behavioural scenario to be successfully classified. If an LLM failed to produce a valid response, the scenario was recorded as an unsuccessful classification rather than being interpreted as a prediction of Normal. For ground-truth Anomaly scenarios, an invalid response therefore contributed to the false negative count used for the end-to-end evaluation. Response validity was also recorded separately, allowing classification performance among valid responses to be distinguished from the overall reliability of the complete LLM classification pipeline. This distinction was particularly important when evaluating models that did not consistently satisfy the structured-response requirements.

Detection performance was evaluated using accuracy, precision, recall, F1-score, false positive rate (FPR), false negative rate (FNR), and Matthews Correlation Coefficient (MCC) (Powers, 2011; Chicco &

Jurman, 2020). Accuracy measured the proportion of correctly classified scenarios across the complete evaluation dataset:

$$\text{Accuracy } = \frac{\text{TP} + \text{TN}}{\text{TP} + \text{TN} + \text{FP} + \text{FN}}$$

Precision measured the proportion of scenarios classified as anomalous that were correctly identified:

$$\text{Precision } = \frac{\text{TP}}{\text{TP} + \text{FP}}$$

Recall measured the proportion of ground-truth anomalous scenarios that were successfully detected:

$$\text{Recall } = \frac{\text{TP}}{\text{TP} + \text{FN}}$$

The F1-score provided the harmonic mean of precision and recall, allowing both false-positive and false-negative behaviour to contribute to the performance measure:

$$F1 = \frac{2 \times (Precision \times Recall)}{Precision + Recall}$$

The false positive rate measured the proportion of ground-truth Normal scenarios incorrectly classified as Anomaly:

$$FPR = \frac{FP}{FP + TN}$$

The false negative rate measured the proportion of ground-truth Anomaly scenarios incorrectly classified as Normal:

$$FNR = \frac{FN}{FN + TP}$$

Matthews Correlation Coefficient was also calculated to provide a balanced measure incorporating all four confusion-matrix outcomes. MCC ranges from -1 to +1, where +1 represents perfect classification, 0 indicates performance with no better correlation than random prediction, and -1 represents complete disagreement between predicted and ground-truth classifications (Chicco & Jurman, 2020):

$$MCC = \frac{(TP \times TN) - (FP \times FN)}{\sqrt{(TP + FP)(TP + FN)(TN + FP)(TN + FN)}}$$

Particular attention was given to recall and false negative rate because missed anomalous behaviour can have significant operational consequences within cybersecurity monitoring environments (Scarfone and Mell, 2007). Accuracy was therefore not considered in isolation when comparing detection effectiveness. Precision and FPR were used to assess the extent to which improved anomaly detection resulted in additional false alarms, while F1-score and MCC provided broader measures of classification performance across the dataset (Opitz, 2024).

For the LLM-based approaches, operational performance was also evaluated using inference latency and response validity. Inference latency measured the time required for a model to return a response for each behavioural scenario, while response validity measured whether the returned output satisfied the structured-response requirements defined in Section 3.5. These measures allowed the evaluation to consider not only anomaly detection performance but also the reliability and practical operation of each LLM within the automated classification pipeline.

### 3.9 GPT-OSS Structured-Output Reliability Experiment

During the initial LLM evaluation, GPT-OSS 20B demonstrated substantially lower structured-response validity than Meta Llama 3.1 8B Instruct and Qwen 2.5 7B Instruct. Although GPT-OSS was evaluated using the same prompt structure, generation parameters, behavioural scenarios, and validation requirements described in Sections 3.5 and 3.6, a large proportion of its responses did not satisfy the structured-output requirements of the automated evaluation pipeline. This raised the question of whether the observed behaviour represented a consistent limitation of the model or variability between separate evaluation runs.

To investigate this behaviour, the complete GPT-OSS evaluation was repeated two additional times. No changes were made to the experimental configuration between runs. The same behavioural scenarios, ground-truth labels, standardised prompt template, behavioural cues, model configuration, generation parameters, API-based inference method, and response-validation procedures were retained. Maintaining identical experimental conditions allowed the additional runs to evaluate whether GPT-OSS produced consistent results when the original experiment was repeated.

Each additional run was evaluated using the same framework described in Section 3.8. Response validity and inference latency were recorded, while valid classifications were compared against the same ground-truth labels used during the original evaluation. The three GPT-OSS runs were then compared to determine the consistency of structured-response validity and anomaly classification performance across repeated evaluations.

The original GPT-OSS evaluation remained part of the primary comparison with Meta Llama 3.1 8B Instruct and Qwen 2.5 7B Instruct because all three models were evaluated under the same initial experimental conditions. The two additional GPT-OSS runs were treated as a separate repeatability analysis and were not substituted for the original results. This preserved the fairness of the primary cross-model comparison while allowing the reliability of GPT-OSS behaviour to be investigated further.

## 4. Results

This section presents the results of the comparative evaluation between Wazuh rule-based detection, OpenSearch statistical anomaly detection, and the three evaluated instruction-tuned LLMs. Performance was assessed using the evaluation framework described in Section 3.8, with each approach evaluated against the ground-truth behavioural dataset.

The results first examine the overall detection performance of each approach using the confusion-matrix outcomes and associated evaluation metrics. This is followed by a comparative analysis of detection behaviour, including performance across the borderline scenarios where the distinction between Normal and Anomaly was less obvious. Operational performance is then examined using LLM inference latency and structured-response validity. Finally, the additional GPT-OSS repeatability experiment described in Section 3.9 is evaluated separately to determine whether the structured-output behaviour observed during the initial evaluation remained consistent across repeated runs under identical experimental conditions.

### 4.1 Overall Detection Performance

Overall detection performance was evaluated across the complete dataset of 75 behavioural scenarios using the ground-truth classifications and performance metrics defined in Section 3.8. Following application of the binary severity threshold described in Section 3.3, the dataset contained 24 Normal scenarios and 51 Anomaly scenarios. Table 4 presents the confusion-matrix outcomes for Wazuh, OpenSearch Anomaly Detection, and the three evaluated LLMs during the primary experiment.

*Table 4. Confusion-Matrix Results Across the 75 Behavioural Scenarios*

| Detection Approach | TP | TN | FP | FN |
|---|---|---|---|---|
| Meta Llama 3.1 8B Instruct | 45 | 22 | 2 | 6 |
| Qwen 2.5 7B Instruct | 35 | 23 | 1 | 16 |
| GPT-OSS 20B | 16 | 24 | 0 | 35 |
| Wazuh | 16 | 23 | 1 | 35 |
| OpenSearch Anomaly Detection | 13 | 24 | 0 | 38 |

*GPT-OSS produced valid structured responses for 22 of the 75 scenarios during the primary evaluation. For the end-to-end comparison in Table 4, scenarios without a valid GPT-OSS classification were treated as unsuccessful detections according to their ground-truth class. Classification performance calculated only from valid GPT-OSS responses is examined separately in Section 4.5.

Meta Llama 3.1 8B Instruct produced the strongest overall detection performance in the primary evaluation, correctly identifying 45 of the 51 anomalous scenarios while correctly classifying 22 of the 24 Normal scenarios. This resulted in six false negatives and two false positives. Qwen 2.5 7B Instruct detected 35 anomalous scenarios and correctly classified 23 Normal scenarios, producing 16 false negatives and one false positive.

Wazuh detected 16 of the 51 anomalous scenarios, resulting in 35 false negatives, while OpenSearch detected 13 anomalous scenarios and produced 38 false negatives. Both baseline approaches maintained comparatively low false positive counts, with Wazuh producing one false positive and OpenSearch producing none. These results indicate that the principal difference between the detection approaches was their ability to identify ground-truth anomalous behaviours rather than their tendency to incorrectly classify Normal scenarios.

Table 5 presents the performance metrics derived from the confusion-matrix outcomes. Meta Llama achieved the highest end-to-end accuracy at 89.3%, together with a recall of 88.2% and an F1-score of 91.8%. Qwen achieved an accuracy of 77.3% and recall of 68.6%. Wazuh and OpenSearch achieved substantially lower recall values of 31.4% and 25.5%, respectively, reflecting the larger number of anomalous scenarios that remained undetected.

*Table 5. Overall Detection Performance Across the 75 Behavioural Scenarios*

| Detection Approach | Accuracy | Precision | Recall | F1-score | FPR | FNR | MCC |
|---|---|---|---|---|---|---|---|
| Meta Llama 3.1 8B Instruct | 89.3% | 95.7% | 88.2% | 91.8% | 8.3% | 11.8% | 0.771 |
| Qwen 2.5 7B Instruct | 77.3% | 97.2% | 68.6% | 80.5% | 4.2% | 31.4% | 0.602 |
| GPT-OSS 20B* | 53.3% | 100.0% | 31.4% | 47.8% | 0.0% | 68.6% | 0.357 |
| Wazuh | 52.0% | 94.1% | 31.4% | 47.1% | 4.2% | 68.6% | 0.303 |
| OpenSearch AD | 49.3% | 100.0% | 25.5% | 40.6% | 0.0% | 74.5% | 0.314 |

*GPT-OSS end-to-end metrics incorporate the effect of invalid structured responses across the complete 75-scenario evaluation. Metrics calculated only from valid GPT-OSS responses are reported separately to distinguish classification capability from structured-output reliability.

Meta Llama demonstrated the strongest balance between anomaly detection and false alarm control, achieving the highest recall, F1-score, and MCC among the evaluated approaches while maintaining an FPR of 8.3%. Qwen also outperformed both traditional baselines in recall and F1-score, although its false negative rate of 31.4% was substantially higher than Meta Llama's 11.8%.

GPT-OSS requires separate interpretation because its primary limitation during the initial evaluation was structured-output reliability. Only 22 of its 75 responses satisfied the required output-validation criteria. Consequently, its end-to-end performance across the complete dataset was substantially lower than its classification performance among valid responses. When only the 22 valid responses were considered, GPT-OSS correctly classified 21, corresponding to an accuracy of 95.5%, precision of 100%, recall of 94.1%, F1-score of 97.0%, and MCC of 0.886. These conditional results should not be directly compared with the complete 75-scenario results of Meta Llama and Qwen, but they indicate that GPT-OSS could classify behavioural scenarios effectively when a valid structured response was successfully produced.

Overall, the primary evaluation shows that Meta Llama provided the strongest reliable end-to-end performance across the complete behavioural dataset. Qwen also demonstrated improved anomaly detection compared with Wazuh and OpenSearch, particularly through higher recall and lower false negative rates. GPT-OSS demonstrated strong classification capability among its valid responses, but its low structured-response validity prevented this capability from translating into reliable end-to-end performance during the initial experiment. The repeatability of this behaviour is examined separately in the GPT-OSS analysis later in this section.

### 4.2 Comparative Detection Analysis

The comparative results show clear differences in the detection behaviour of the evaluated approaches. Meta Llama 3.1 8B Instruct demonstrated the strongest balance between detecting anomalous behaviour and limiting false classifications. Of the 51 ground-truth Anomaly scenarios, Meta Llama correctly detected 45, resulting in a recall of 0.882 and a false negative rate of 0.118. Qwen 2.5 7B Instruct detected 35 anomalous scenarios, producing a lower recall of 0.686 and a false negative rate of 0.314. Both models therefore detected a greater proportion of anomalous behaviours than the Wazuh and OpenSearch baselines.

The difference was most apparent in the number of missed anomalies. Wazuh correctly detected 16 of the 51 anomalous scenarios and failed to identify 35, resulting in a recall of 0.314 and false negative rate of 0.686. OpenSearch detected 13 anomalous scenarios and missed 38, producing the lowest recall of 0.255 and highest false negative rate of 0.745. In comparison, Meta Llama reduced the number of false negatives to six, while Qwen reduced the number to 16.

The increased recall of the LLM-based approaches did not result in a substantial increase in false positive detections. Meta Llama incorrectly classified two of the 24 Normal scenarios as Anomaly, corresponding to an FPR of 0.083, while Qwen produced one false positive and an FPR of 0.042. Wazuh also produced one false positive, while OpenSearch produced none. OpenSearch therefore achieved perfect precision and a zero false positive rate, but this occurred alongside substantially lower anomaly detection, with 38 of the 51 anomalous scenarios remaining undetected.

The MCC results further demonstrate the differences in overall classification performance. Meta Llama achieved the highest MCC of 0.771, followed by Qwen at 0.602. OpenSearch and Wazuh achieved MCC values of 0.314 and 0.303, respectively. These results indicate that Meta Llama provided the strongest overall agreement with the ground-truth classifications across both Normal and Anomaly scenarios.

GPT-OSS 20B presented a different performance profile from the other LLMs. Its end-to-end results across the 75 scenarios produced a recall of 0.314 and an F1-score of 0.478. However, as identified in Section 4.1, these results were strongly affected by the model's low structured-response validity rather than classification errors alone. Consequently, GPT-OSS cannot be interpreted in the same manner as Meta

Llama and Qwen without also considering its response-validity behaviour. This operational characteristic is examined further in Sections 4.4 and 4.5.

Overall, the comparative results show that Meta Llama and Qwen identified a greater proportion of ground-truth anomalous behaviours than Wazuh and OpenSearch while maintaining relatively low false positive rates. The largest performance difference between the approaches was observed in false negative behaviour, providing the basis for examining detection performance on the more difficult borderline scenarios in Section 4.3.

### 4.3 Borderline Behaviour Detection

The overall results demonstrate differences in anomaly detection performance across the evaluated approaches; however, the borderline scenarios provide a more focused assessment of behaviours where the distinction between Normal and Anomaly was less obvious. As described in Sections 3.2 and 3.3, the dataset contained 24 scenarios categorised as borderline or suspicious. Four of these scenarios were assigned a ground-truth severity score of 1 and were therefore classified as Normal, while 20 were assigned severity 2 and classified as Anomaly under the predefined binary decision threshold.

The 20 severity-2 borderline scenarios were analysed separately to determine how effectively each detection approach identified suspicious behaviours requiring investigation. These scenarios represented less obvious anomalous activity than the higher-severity cases and therefore provided a useful basis for examining differences between rule-based, statistical, and LLM-based detection.

*Table 6. Detection Performance Across the 20 Severity-2 Borderline Anomaly Scenarios*

| ***Detection Approach*** | ***Detected as Anomaly*** | ***Missed/Not Detected*** | ***Detection Rate*** |
|---|---|---|---|
| *Meta Llama 3.1 8B Instruct* | *16* | *4* | *80.0%* |
| Qwen 2.5 7B Instruct | *11* | *9* | *55.0%* |
| *GPT-OSS 20B** | *5* | *15* | *25.0%* |
| *Wazuh Rule-Based Detection* | *4* | *16* | *20.0%* |
| *OpenSearch AD* | *3* | *17* | *15.0%* |

*Note: GPT-OSS produced valid structured responses for only 5 of the 20 severity-2 borderline scenarios during the primary evaluation. All five valid responses correctly classified the corresponding scenarios as Anomaly. The remaining 15 scenarios did not produce valid structured outputs and therefore should not be interpreted as classifications of Normal. Consequently, the 25.0% detection rate shown for GPT-OSS reflects end-to-end pipeline performance rather than its classification accuracy among valid responses.*

Meta Llama and Qwen can then be compared with the traditional detection baselines based on the proportion of severity-2 borderline scenarios correctly identified as Anomaly. Particular attention should be given to scenarios detected by the LLMs but missed by Wazuh or OpenSearch, as these cases provide evidence of differences in detection behaviour across the three approaches.

The borderline analysis should also be interpreted alongside the false-positive results for the four severity-1 borderline scenarios classified as Normal. This distinction is important because improved detection of severity-2 behaviours should not be considered in isolation from the possibility of incorrectly escalating plausible severity-1 activity. Together, the severity-1 and severity-2 borderline cases provide a more detailed assessment of detection behaviour around the study's Normal/Anomaly decision threshold.

### 4.4 Operational Performance

In addition to classification performance, the operational performance of the three LLMs was evaluated using inference latency and structured-response validity. These measures were included because an LLM-based detection approach must not only classify behavioural scenarios accurately but also produce responses that can be reliably processed within an automated security monitoring workflow.

Table 7 presents the average inference latency and structured-response validity observed during the primary 75-scenario evaluation.

*Table 7. Operational Performance of the Evaluated LLMs*

| Model | Average Latency (s) | Valid Responses | Invalid Responses | Response Validity |
|---|---|---|---|---|
| Meta Llama 3.1 8B Instruct | 1.86 | 75 | 0 | 100.0% |
| Qwen 2.5 7B Instruct | 0.83 | 75 | 0 | 100.0% |
| GPT-OSS 20B | 4.94 | 22 | 53 | 29.3% |

Meta Llama 3.1 8B Instruct and Qwen 2.5 7B Instruct produced valid structured responses for all 75 behavioural scenarios, resulting in a response-validity rate of 100%. Qwen also recorded the lowest average inference latency at 0.83 seconds, compared with 1.86 seconds for Meta Llama. Both models therefore demonstrated consistent compliance with the structured-response requirements of the automated evaluation pipeline.

GPT-OSS 20B exhibited substantially different operational behaviour during the primary evaluation. The model produced only 22 valid structured responses from the 75 behavioural scenarios, resulting in a response-validity rate of 29.3%. The remaining 53 responses failed the validation requirements described in Section 3.5. GPT-OSS also recorded the highest average inference latency at 4.94 seconds.

The low response-validity rate is particularly important when interpreting the classification performance of GPT-OSS. As reported in Section 4.1, GPT-OSS demonstrated strong classification performance among the responses that satisfied the validation requirements, correctly classifying 21 of its 22 valid responses. However, the high number of invalid responses substantially reduced its end-to-end reliability across the complete evaluation dataset. This demonstrates a distinction between classification performance on valid outputs and the ability of a model to consistently produce structured responses suitable for automated processing.

The operational results therefore show that Meta Llama and Qwen provided substantially more consistent structured-output behaviour than GPT-OSS during the primary evaluation. Qwen provided the lowest inference latency, while Meta Llama combined 100% response validity with the strongest overall anomaly detection performance reported in Section 4.1. The repeatability of GPT-OSS's structured-output behaviour was investigated through the two additional evaluation runs described in Section 3.9, with the results presented in Section 4.5.

### 4.5 GPT-OSS Repeatability Analysis

The primary evaluation identified substantially lower structured-response validity for GPT-OSS 20B than for Meta Llama 3.1 8B Instruct and Qwen 2.5 7B Instruct. As described in Section 3.9, the complete GPT-OSS evaluation was therefore repeated twice under the same experimental conditions to determine whether the structured-output behaviour observed during the initial run remained consistent across repeated evaluations.

Table 8 compares the operational and classification results from the three GPT-OSS evaluation runs. No experimental parameters were intentionally changed between runs, and each evaluation used the same 75 behavioural scenarios, ground-truth classifications, prompt structure, generation settings, and response-validation requirements.

*Table 8. GPT-OSS 20B Performance Across Three Evaluation Runs*

| GPT-OSS Run | Valid Responses | Invalid Responses | Response Validity | Valid-Response Accuracy | Valid-Response Recall | Valid-Response F1 Score |
|---|---|---|---|---|---|---|
| Run 1 | 22 | 53 | 29.3% | 95.5% | 94.1% | 97.0% |
| Run 2 | 29 | 46 | 38.7% | 89.7% | 84.2% | 91.4% |
| Run 3 | 29 | 46 | 38.7% | 86.2% | 81.8% | 90.0% |

Structured-response validity remained low across all three evaluations. The initial run produced 22 valid responses from the 75 scenarios, corresponding to a validity rate of 29.3%. Runs 2 and 3 each produced 29 valid responses, increasing the validity rate to 38.7%. Despite this increase, more than 60% of GPT-OSS responses remained invalid in both additional runs.

Classification performance among the valid responses remained comparatively high across all three runs. Run 1 achieved a valid-response accuracy of 95.5%, recall of 94.1%, and F1-score of 97.0%. Run 2 achieved an accuracy of 89.7%, recall of 84.2%, and F1-score of 91.4%, while Run 3 achieved an accuracy of 86.2%, recall of 81.8%, and F1-score of 90.0%. These results show that GPT-OSS generally classified behavioural scenarios successfully when it produced a response that satisfied the validation requirements, although classification performance among valid responses also varied between runs.

The repeated evaluations therefore produced a consistent distinction between classification capability and structured-output reliability. GPT-OSS demonstrated strong classification performance among valid responses, but valid structured outputs were produced for fewer than 40% of the 75 scenarios in each of the three runs. The additional evaluations therefore confirmed that the low response-validity rate observed during the primary experiment was not limited to a single evaluation run.

No changes were intentionally introduced between the three evaluations. Consequently, the variation observed across runs cannot be attributed within this experiment to changes in the dataset, prompt structure, ground-truth labels, or configured generation parameters. The results demonstrate variability under repeated evaluation conditions; however, the experiment does not establish the underlying cause of that variability.

The original GPT-OSS results were retained for the primary cross-model comparison presented in Sections 4.1–4.4, while Runs 2 and 3 were used only to assess repeatability. This ensured that the additional evaluations did not replace the original result or alter the conditions used to compare GPT-OSS with Meta Llama and Qwen.

## 5. Discussion and Conclusion

This research contributes to cybersecurity anomaly detection in three key areas. First, a standardised instruction-based LLM classification framework was developed to evaluate authentication behaviours using a defined anomaly severity scale, behavioural cues, and structured-response requirements. Second, a controlled cybersecurity testbed was developed to generate endpoint-specific authentication data containing normal, borderline, and anomalous behavioural scenarios. Third, three recent instruction-tuned LLMs, Meta Llama 3.1 8B Instruct, Qwen 2.5 7B Instruct, and GPT-OSS 20B, were evaluated against Wazuh rule-based detection and OpenSearch statistical anomaly detection using the behavioural dataset generated from the testbed.

Detection performance was evaluated using accuracy, precision, recall, F1-score, false positive rate, false negative rate, and Matthews Correlation Coefficient. The LLMs were also evaluated using inference latency and structured-response validity. Meta Llama 3.1 8B Instruct achieved the strongest overall detection performance, with an accuracy of 89.3%, recall of 88.2%, F1-score of 91.8%, and false negative rate of 11.8%. Qwen 2.5 7B Instruct achieved an accuracy of 77.3% and recall of 68.6%. In comparison, Wazuh achieved a recall of 31.4% and false negative rate of 68.6%, while OpenSearch achieved a recall of 25.5% and false negative rate of 74.5%.

The main weakness observed with Wazuh and OpenSearch was not the number of false positives, but the number of anomalous scenarios they failed to detect. Both approaches maintained low false positive rates, with Wazuh at 4.2% and OpenSearch at 0%. However, Wazuh failed to detect 35 of the 51 anomalous scenarios, while OpenSearch failed to detect 38. In comparison, Meta Llama produced only six false negatives while maintaining a low number of false positives at two.

A similar result was observed when only the 20 severity-2 borderline anomalous scenarios were evaluated. Meta Llama detected 16 of the 20 scenarios, while Qwen detected 11. Wazuh detected four and OpenSearch detected three. This follows the same general pattern observed across the complete dataset, with Meta Llama achieving the highest detection rate, followed by Qwen, Wazuh, and OpenSearch. These borderline scenarios were important to the evaluation because they were the suspicious cases closest to the Normal/Anomaly classification threshold.

The operational results also showed important differences between the three LLMs. Qwen achieved the lowest average inference latency at 0.83 seconds, followed by Meta Llama at 1.86 seconds and GPT-OSS at 4.94 seconds during the primary evaluation. Meta Llama and Qwen produced valid structured responses for all 75 scenarios. GPT-OSS, however, produced only 22 valid responses, giving a structured-response validity rate of 29.3%. Although GPT-OSS performed well when it produced a valid response, the large number of invalid responses reduced its overall reliability within the evaluation pipeline. The complete GPT-OSS evaluation was therefore repeated twice using the same experimental settings. Both additional runs produced 29 valid responses from the 75 scenarios, giving a validity rate of 38.7%. This showed that the structured-output problem was not limited to the original evaluation.

Overall, the results show that instruction-tuned LLMs can provide an additional analysis layer for authentication anomaly detection. Meta Llama provided the strongest combination of anomaly detection and reliable structured responses, while Qwen provided the fastest inference time with 100% response validity. GPT-OSS also showed that classification performance is only one part of evaluating an LLM for an automated cybersecurity workflow. A model must also produce consistent structured responses that can be reliably processed by the surrounding detection pipeline.

### 5.1 Why Traditional Detection Falls Short

The main limitation observed with the traditional detection approaches was their ability to identify anomalous behaviour. Wazuh and OpenSearch maintained low false positive rates of 4.2% and 0%, respectively, but produced much higher false negative rates of 68.6% and 74.5%. Of the 51 ground-truth anomalous scenarios, Wazuh detected 16 and OpenSearch detected 13, leaving 35 and 38 scenarios undetected.

These results can partly be explained by how the two approaches detect suspicious activity. Wazuh relies on predefined rules and conditions to identify security events. This worked well when authentication activity matched an existing rule, as demonstrated by the alerts generated during the experiment. However, suspicious behaviour can involve several events that may appear normal or less concerning when viewed individually. If the overall behaviour does not meet an existing rule condition, the individual events may still be recorded without the broader pattern being identified as anomalous (Agarwal & Hussain, 2018; Guide et al., 2024).

OpenSearch takes a different approach by using statistical anomaly detection to identify changes from previously observed behaviour. This allows unusual activity to be detected without requiring a predefined attack signature. However, a statistical change does not always mean that an event is a security threat, just as suspicious behaviour may not always create a large enough statistical change to be detected. This can become particularly important for less obvious activity that remains close to the expected behavioural baseline (Yeom & Jung, 2022; Zha et al., 2023).

This was also seen in the severity-2 borderline scenarios. Wazuh detected four of the 20 borderline anomalies, giving a detection rate of 20%, while OpenSearch detected three, giving a detection rate of 15%. These results were lower than their overall anomaly recall of 31.4% and 25.5%, showing that both approaches had greater difficulty detecting the less obvious borderline behaviours.

The results do not suggest that Wazuh or OpenSearch should be replaced. Both approaches provided useful detection capabilities and maintained low false positive rates during the experiment. Instead, the high number of false negatives shows a limitation when either approach is used on its own to detect behaviours that may require additional context. This provides a reason to investigate an additional detection layer that can consider the relationship between authentication events alongside the existing rule-based and statistical approaches.

### 5.2 LLM-Based Semantic Detection and Improved Recall

Meta Llama 3.1 8B Instruct and Qwen 2.5 7B Instruct achieved higher anomaly recall than Wazuh and OpenSearch. Meta Llama achieved a recall of 88.2%, detecting 45 of the 51 ground-truth anomalous scenarios, while Qwen achieved a recall of 68.6%, detecting 35. In comparison, Wazuh achieved a recall of 31.4% and OpenSearch achieved 25.5%. This difference can also be seen in the number of false negatives. Meta Llama produced six false negatives and Qwen produced 16, compared with 35 for Wazuh and 38 for OpenSearch.

The higher recall achieved by Meta Llama and Qwen did not result in a large increase in false positives. Meta Llama produced two false positives across the 24 Normal scenarios, while Qwen and Wazuh each produced one and OpenSearch produced none. Meta Llama therefore detected 29 more anomalous scenarios than Wazuh while producing only one additional false positive. This is important in cybersecurity monitoring because improving anomaly detection is less useful if it also creates a large number of unnecessary alerts for security analysts.

A similar pattern was observed in the 20 severity-2 borderline anomalous scenarios. Meta Llama detected 16 of the 20 scenarios, giving a detection rate of 80%, while Qwen detected 11, giving a detection rate of 55%. Wazuh detected four and OpenSearch detected three. Although detection rates were lower for these less obvious scenarios, the overall order remained the same, with Meta Llama achieving the highest detection rate, followed by Qwen, Wazuh, and OpenSearch. This shows that the difference observed across the complete dataset was also present in the scenarios closest to the Normal/Anomaly threshold.

One possible reason for this difference is how the authentication behaviour was presented to the LLMs. Rather than analysing individual authentication events, the models received natural-language descriptions containing the broader context of each behavioural scenario. The standardised prompt also provided the seven behavioural cues described in Section 3.6, including timing variance, repetitive sequences, impossible travel, rotating IP addresses, and extremely fast interactions. This gave the models additional context when deciding whether a scenario should be classified as Normal or Anomaly. Prior work likewise identifies contextualisation and semantic interpretation as important for security analysis (Eckhoff et al., 2025; Singh et al., 2025). The results support further investigation of semantic analysis as an additional method for detecting suspicious authentication behaviour. However, this experiment does not prove that semantic reasoning alone caused the improved detection performance.

The results also support using LLM-based classification as an additional analysis layer rather than replacing existing SIEM detection methods. Wazuh and OpenSearch already provide useful rule-based and statistical detection capabilities. An LLM-based layer could provide additional analysis when the behaviour is difficult to classify using predefined rules or statistical changes alone. This human-in-the-loop, cognitive-aid positioning is consistent with empirical observations of LLM use in SOC work (Singh et al., 2025). This would allow the different approaches to work together rather than requiring one detection method to replace another.

### 5.3 Operational Implications

Detection performance is important, but an LLM used within a cybersecurity monitoring environment also needs to produce consistent responses within a reasonable amount of time. This is particularly important for automated workflows, where the output from the model needs to be processed by another system. For this reason, inference latency and structured-response validity were considered alongside the classification results.

Qwen 2.5 7B Instruct achieved the lowest average inference latency at 0.83 seconds per behavioural scenario, followed by Meta Llama 3.1 8B Instruct at 1.86 seconds. Both models also produced valid structured responses for all 75 scenarios, achieving 100% response validity. This shows that both models were able to consistently follow the structured-output requirements of the classification pipeline. Qwen provided the fastest response time, while Meta Llama achieved the strongest overall detection performance.

GPT-OSS 20B produced a different result. During the primary evaluation, it produced valid structured responses for only 22 of the 75 scenarios, giving a response-validity rate of 29.3%. It also recorded the highest average inference latency at 4.94 seconds. However, 21 of its 22 valid responses were classified correctly. This shows that GPT-OSS could perform well when it returned a valid response, but the large number of invalid responses made it less reliable within the complete evaluation pipeline.

The two additional GPT-OSS evaluations showed a similar problem. Response validity increased to 38.7% in both runs but remained below 40%. The scenarios that produced valid responses also changed between the three evaluations, and the classification results among those valid responses were not identical. This shows that the structured-output problem was not limited to the first evaluation. However, the experiment cannot determine why this behaviour occurred.

These results show why classification performance should not be considered on its own when selecting an LLM for an automated cybersecurity workflow. Even if a model performs well when a valid response is produced, it may still create problems if its output cannot be consistently processed by the surrounding system. Response validity, inference latency, error handling, and output validation are therefore also important. Invalid responses may require additional handling through validation, retries, or another detection method when a usable response cannot be produced. More broadly, instruction-following reliability varies across models and tasks, so adherence to output requirements should be evaluated rather than assumed (Lou et al., 2024).

Overall, the three LLMs showed different strengths and limitations. Meta Llama provided the strongest combination of anomaly detection and reliable structured responses. Qwen was the fastest model and achieved 100% response validity, although its anomaly recall was lower than Meta Llama. GPT-OSS performed well across many of its valid responses but had much lower structured-response reliability and the highest inference latency. These results show that selecting an LLM for cybersecurity monitoring requires consideration of both its detection performance and how reliably it can operate within the wider system.

### 5.4 Theoretical Implications

The results of this study provide some important considerations for using LLMs in cybersecurity anomaly detection. Rule-based and statistical approaches identify suspicious behaviour in different ways. Wazuh uses predefined rules and conditions, while OpenSearch identifies statistical changes from expected behaviour. The LLM-based approach adds another method by considering several characteristics of an authentication scenario together before deciding whether the behaviour should be classified as Normal or Anomaly.

Representing authentication activity as behavioural scenarios allowed the LLMs to consider more context than would be available from individual events viewed separately. This was particularly noticeable in the severity-2 borderline scenarios. Meta Llama detected 80% of these scenarios and Qwen detected 55%, compared with 20% for Wazuh and 15% for OpenSearch. These scenarios were directly above the Normal/Anomaly threshold and represented suspicious behaviours that were less obvious than the higher-severity cases. The same general performance order was still observed within this smaller subset, supporting further research into the use of contextual behavioural information for anomaly detection.

The results also show that LLM-based semantic analysis does not need to replace existing rule-based or statistical detection. Each approach considers suspicious activity differently. Wazuh checks activity against predefined rules, OpenSearch identifies statistical changes in behaviour, and the LLMs evaluate a natural-language description containing the broader context of the authentication scenario. Using these approaches together could provide different ways of assessing the same security activity and help identify behaviours that may be difficult for one detection method to recognise on its own.

The instruction-based classification framework is also an important part of the study. Each LLM received the same severity scale, classification requirements, seven behavioural cues, and structured-output requirements. This allowed the same cybersecurity classification task to be given to each model without changing or fine-tuning the underlying models. The results show that existing instruction-tuned LLMs can be used for a specific cybersecurity classification task when they are provided with consistent instructions and relevant behavioural information. However, this study did not compare different prompt designs or evaluate the models without the seven behavioural cues. The results therefore show that the approach is feasible, but they do not show that the prompt used in this experiment is the best possible design.

Finally, the GPT-OSS results show that classification performance and operational reliability need to be considered separately. GPT-OSS performed well across many of the valid responses it produced, but it frequently failed to return a response that satisfied the structured-output requirements. This is important for future LLM-based cybersecurity research because classification metrics alone may not show whether a model can operate reliably within an automated detection pipeline. Future evaluations should therefore consider both how accurately a model classifies security behaviour and how consistently its output can be processed by the surrounding system.

### 5.5 Threats to Validity and Limitations

Several limitations need to be considered when interpreting the results of this study. First, the evaluation was conducted using a controlled testbed and a dataset of 75 behavioural scenarios. The testbed allowed normal, borderline, and anomalous authentication behaviours to be generated and evaluated under consistent conditions. However, it does not reproduce the scale, background traffic, diversity, or complexity of a real enterprise environment. The results should therefore be considered within the controlled environment used for this experiment and should not be assumed to represent performance across all cybersecurity environments.

Second, the dataset intentionally contained a higher proportion of suspicious and anomalous behaviour than would normally be expected within real-world network traffic. After applying the ground-truth severity threshold, the dataset contained 24 Normal and 51 Anomaly scenarios. This distribution was useful for evaluating anomaly detection and false negatives, but it may also influence measures such as accuracy and does not represent the lower proportion of malicious activity normally expected in operational environments. For this reason, precision, recall, F1-score, FPR, FNR, and MCC were evaluated alongside accuracy rather than relying on accuracy alone (Opitz, 2024).

Third, the ground-truth severity labels were assigned using the criteria defined for this experiment. The same severity scale and Normal/Anomaly threshold were used consistently throughout the evaluation, but assigning severity to behavioural activity can still involve judgement, particularly for borderline scenarios. The difference between severity 1 and severity 2 is especially important because it changes the binary classification from Normal to Anomaly. Future research could improve this process by having multiple cybersecurity analysts or experts independently assess the scenarios and compare their classifications.

Fourth, the results may also be influenced by the standardised prompt and the seven behavioural cues provided to the LLMs. The same prompt was used for each model to keep the evaluation consistent, but this study did not compare different prompt designs or repeat the experiment without the behavioural cues. Therefore, although Meta Llama and Qwen achieved higher anomaly recall than Wazuh and OpenSearch, the experiment does not prove that semantic reasoning alone caused the improvement. It also does not show that the prompt used in this study was the best possible configuration.

Fifth, the LLM evaluation was performed through the Hugging Face Router rather than using locally hosted models. This meant that parts of the inference environment, including the serving infrastructure, were outside the direct control of the experiment. This is particularly important for GPT-OSS, where structured-response validity changed across the three evaluations even though no intentional changes were made to the experimental settings. The repeated evaluations confirmed that the structured-output problem was not limited to the first run, but this experiment cannot determine whether the cause was the model, inference provider, serving configuration, or another part of the hosted inference process.

Finally, the experiment focused specifically on authentication-related behavioural scenarios generated within the developed testbed. The results therefore cannot yet be applied to all types of cybersecurity activity. Other attack categories, such as malware activity, network intrusion, privilege escalation, lateral movement, and data exfiltration, were not evaluated. Further testing using larger and more diverse datasets, additional endpoint types, and other attack scenarios is needed to determine whether similar results can be achieved across broader cybersecurity environments.

### 5.6 Future Research

Future research should expand the testbed and behavioural dataset used in this study. The current experiment focused on authentication-related activity, but the testbed could also be used to generate other cybersecurity behaviours, including privilege escalation, lateral movement, network intrusion, malware activity, and data exfiltration. Testing these additional attack categories would help determine whether the improvements observed for authentication anomaly detection can also be achieved across other types of cybersecurity activity.

The ground-truth labelling process could also be improved by having multiple cybersecurity analysts or experts independently assess the behavioural scenarios. Their assigned severity scores could then be compared to determine how consistently the scenarios are classified. This would be particularly useful for behaviours around the severity-1 and severity-2 threshold, where the difference between normal and anomaly is less obvious.

Further testing should also examine how the prompt and seven behavioural cues affected the LLM results. An ablation study could repeat the experiment using different versions of the prompt, such as removing the seven behavioural cues, providing less contextual information, using raw or minimally processed log information, or changing the severity instructions. Comparing these results would help determine how much the behavioural context and prompt design contributed to the improved anomaly detection observed in this study.

Another area for future research is locally hosted LLM inference. Running the models on controlled hardware would provide greater control over the model versions and inference environment and reduce the dependence on external API infrastructure. This would be particularly useful for investigating the structured-output problems observed with GPT-OSS 20B. Repeating the GPT-OSS experiment using a locally hosted model could help determine whether the same response-validity problems occur when the inference environment (e.g., internet connectivity, token limits, and API formatting behaviour) is under direct control.

The LLM classification pipeline could also be integrated with Wazuh or OpenSearch for real-time analysis. Instead of evaluating previously generated behavioural scenarios, authentication activity could be collected directly from the monitoring environment and grouped into behavioural context over a defined period. Selected activity could then be submitted to an LLM for additional analysis. This would allow Wazuh and OpenSearch to continue providing their existing detection capabilities while using the LLM as an additional layer when more behavioural context is required.

Finally, the approach should be evaluated using larger datasets and longer experiments that more closely represent real network environments. Future testbeds could include more endpoint types, multiple users, IoT devices, increased background traffic, and longer behavioural sequences. Testing under these conditions would help determine whether LLM-based analysis can continue to reduce false negatives while maintaining acceptable false positive rates, inference latency, and response reliability as the environment becomes larger and more complex.

Overall, this study shows the potential for instruction-tuned LLMs to complement existing cybersecurity detection methods by providing an additional layer of analysis for authentication behaviour. Meta Llama 3.1 8B Instruct achieved the strongest overall detection performance, reducing false negatives compared with Wazuh and OpenSearch while maintaining a low false positive rate. Qwen 2.5 7B Instruct also improved anomaly detection while providing the lowest inference latency and 100% structured-response validity. The GPT-OSS results showed that classification performance alone is not enough when considering an LLM for an automated cybersecurity workflow, as the model must also produce responses that can be accurately validated. Rather than replacing existing rule-based and statistical detection systems, the results support further research into using LLMs as an additional semantic analysis layer to help turn complex security data from noise into useful signals.